\documentclass[aps,prl,twocolumn,groupedaddress]{revtex4-1}
\usepackage[latin9]{inputenc}
\usepackage{verbatim}
\usepackage{graphicx}
\usepackage{xargs}[2008/03/08]
\usepackage{times}
\usepackage{amsmath}
\usepackage{geometry}
\usepackage{braket}
\usepackage{mathtools}
\usepackage{mathrsfs}
\usepackage{tensor}
\usepackage{geometry}
\usepackage{bbold}
\newcommand{\modulus}[1]{\left|#1\right|}
\def\d{\mbox{d}}
\def\ud{\textrm{d}}
\def\w{\omega}

\begin{document}

	\preprint{}

\title{Single-photon characterization by two-photon spectral interferometry}

\author{Val\'{e}rian Thiel{$^{*1,2}$}}
\author{Alex O. C. Davis{$^{1,3}$}}
\author{Ke Sun{$^{4}$}}
\author{Peru D'Ornellas{$^{1}$}}
\author{Xian-Min Jin{$^{4}$}}
\author{Brian J. Smith{$^{1,2}$}}

\affiliation{{$^1$}Clarendon Laboratory, University of Oxford, Parks Road, Oxford, OX1 3PU, UK}
\affiliation{{$^2$}Department of Physics and Oregon Center for Optical, Molecular, and Quantum Science, University of Oregon, Eugene, Oregon 97403, USA }
\affiliation{{$^3$}Laboratoire Kastler-Brossel, UPMC-Sorbonne Universit\'{e}s, CNRS, ENS-PSL  Research University, Coll\'{e}ge de France, 75005 Paris, France}
\affiliation{{$^4$}Department of Physics, Shanghai Jiao Tong University, Shanghai 200240, PR China}

\affiliation{{$^*$}Corresponding author: vthiel@uoregon.edu}
	
	\begin{abstract}
		Single-photon sources are a fundamental resource in quantum optics. The indistinguishability and purity of photons emitted from different sources are crucial (necessary, essential) properties for many quantum applications to ensure high-visibility interference between different sources. The state of a single-photon source is described by the modes occupied by the single light quanta. Thus the ability to determine the mode structure of a single-photon source provides a means to assess its quality, compare different sources, and provide feedback for source engineering. Here, we propose and demonstrate an experimental scheme that allows for complete characterization of the spectral-temporal state of a pulsed single-photon source. The density matrix elements of the single-photon source are determined by spectral interferometry with a known single-photon reference. Frequency-resolved coincidence measurements are performed after the unknown single-photon source is interfered with a single-photon reference pulse. Fourier analysis of the frequency-resolved two-photon interference pattern reveals the spectral-temporal density matrix of the broadband single-photon source. We present an experimental realization of this method for pure and mixed state pulsed, single-photon sources.
	\end{abstract}
	
	\date{\today}
	
	\maketitle

\section{Introduction}

Two-photon interference, in which indistinguishable photons are incident on the two input ports of a balanced beam splitter and always emerge from the same output port when the photons are identical and pure, is a foundational aspect of the quantum nature of light. First demonstrated by Hong, Ou and Mandel \cite{hong:87}, this nonclassical effect reflects the bosonic nature of light and is at the heart several quantum information protocols including boson sampling \cite{spring2013boson,broome2013photonic,tillmann2013experimental}, linear optics quantum computing \cite{kok2007linear}, and quantum networks that utilise quantum repeaters \cite{kimble2008quantum}. The suppression of single-photon events at each output of the beam splitter, which provides the visibility in Hong-Ou-Mandel Interference (HOMI), depends critically on the indistinguishability of the photons and their purity \cite{mosley:08,PRLsun}. A means to determine the state of single-photon sources is thus essential to diagnosing and engineering improved source design and matching. 

The state of a single photon is given by the mode distribution occupied by the light quantum. In general, this can be represented by a density matrix. The mode structure of light arises from solutions to the classical Maxwell equations which describe the polarization, transverse-spatial, and spectral-temporal degrees of freedom (DOF) for a beam of light \cite{birula:94,smith:07}. Here we focus on the spectral-temporal degree of freedom, which has gained significant interest recently for quantum information applications \cite{brecht:15,nunn:13}. 

Characterization of single-photon pulse mode structure has been demonstrated by interference with a known reference field and more recently in a self-referencing interferometer. The self-referencing method has demonstrated reconstruction of single-mode, pure-state, photon sources, but has yet to achieve characterisation of partially coherent (partially mixed) single-photon states. While the methods using reference pulses, which utilise either a strong classical field \cite{lvovsky:09b, bellini:04} to implement a mode-selective projective measurement, or a weak single-photon-level field \cite{wasilewski:07}, have shown reconstruction of mixed state single-photon sources, these require the reference to be scanned. Scanning the reference field introduces experimental challenges related to stability and time required for data acquisition.

Here we propose and demonstrate a method to determine the mixed pulse-mode state of a single-photon source by interfering the single-photon source with a reference single-photon source at a balanced beam splitter and performing frequency-resolved coincidence measurements at the output, frequency-resolved HOMI (FR-HOMI). Fourier analysis of the FR-HOMI data enables one to directly determine the density matrix of the unknown source when the reference occupies a single pulse mode that is spectrally broader than it. This method requires only one experimental configuration and does not require scanning of the reference pulse making it fast and resilient to experimental fluctuations.

\section{Single-photon state and frequency-resolved HOMI}
\label{2dAlgorithm}
The state of a spectrally-pure single photon is represented by the complex-valued spectral-mode function, $\tilde{\psi}(\w)$, which may be interpreted as the spectral wave function \cite{smith:07}. The state of the electromagnetic field when a single photon occupies the mode $\tilde{\psi}(\omega)$ can be expressed by $\Ket{1}_{\psi} = \hat{a}^\dagger_\psi\Ket{\mbox{vac}}$, where $\Ket{\mbox{vac}}$ is the vacuum state of the field and
\begin{equation}
\label{1photonstate}
\hat{a}^\dagger_\psi= \int \textrm{d}\omega~ \tilde{\psi}(\omega) {\hat{a}}^{\dagger}(\omega),
\end{equation}
\noindent where ${\hat{a}}^{\dagger}(\omega)$ is a field operator creating a single monochromatic photon of frequency $\omega$. In general, a single-photon source can emit a photon that does not occupy one mode, but rather a mixture of modes $\{\tilde{\psi}_i(\omega)\}$. The state of such a source cannot be represented by a single mode function, but rather by the density operator $\hat{\rho} = \sum_i P_i \Ket{1}_{i} {}_{i} \Bra{1}$, where $\Ket{1}_{i}=\hat{a}^\dagger_{i}\Ket{\mbox{vac}}$ is a single-photon state in mode $\{\tilde{\psi}_i(\omega)\}$ and  the real non-negative coefficients $P_i$ satisfy $\sum_i P_i =1$ \cite{fano:57}. The density operator may be equivalently written in the spectral-mode basis as
\begin{equation}
\hat{\rho}=\iint \rho(\w,\w') \Ket{\w}\Bra{\w'} \d\w \d\w',
\end{equation}
where $\rho(\w,\w') = \Braket{\w|\hat{\rho}|\w'}$ is the spectral density matrix element and $\Ket{\w} = \hat{a}^\dagger(\w)\Ket{\mbox{vac}}$ is a monochromatic single-photon state. Hence, full information about the spectral-temporal state of the photon is encapsulated in the spectral density matrix elements $\rho(\w,\w')$.

To determine the density matrix elements in the frequency basis, $\rho(\w,\w')$, we utilize Hong-Ou-Mandel interference (HOMI) between the unknown single-photon source and a known reference single-photon source. When an unknown single-mode single-photon source is interfered with a calibrated single-photon source on a balanced beam splitter, no coincidence detection events should be registered between the outputs for the case in which the reference photon occupies the same mode as the unknown source. It has been experimentally demonstrated that by recording the coincidence count rates while scanning the mode structure of the reference pulse, one can obtain the density matrix elements of the unknown source \cite{wasilewski:07}. In this case the reference was derived from a well-characterised, highly-attenuated shaped laser pulse. Although this technique can characterize the single-photon density matrix elements, it requires repeated reconfiguration of the reference signal, necessitating long acquisition times and stable conditions.
\begin{figure}[t]
	\includegraphics[width=.75\linewidth]{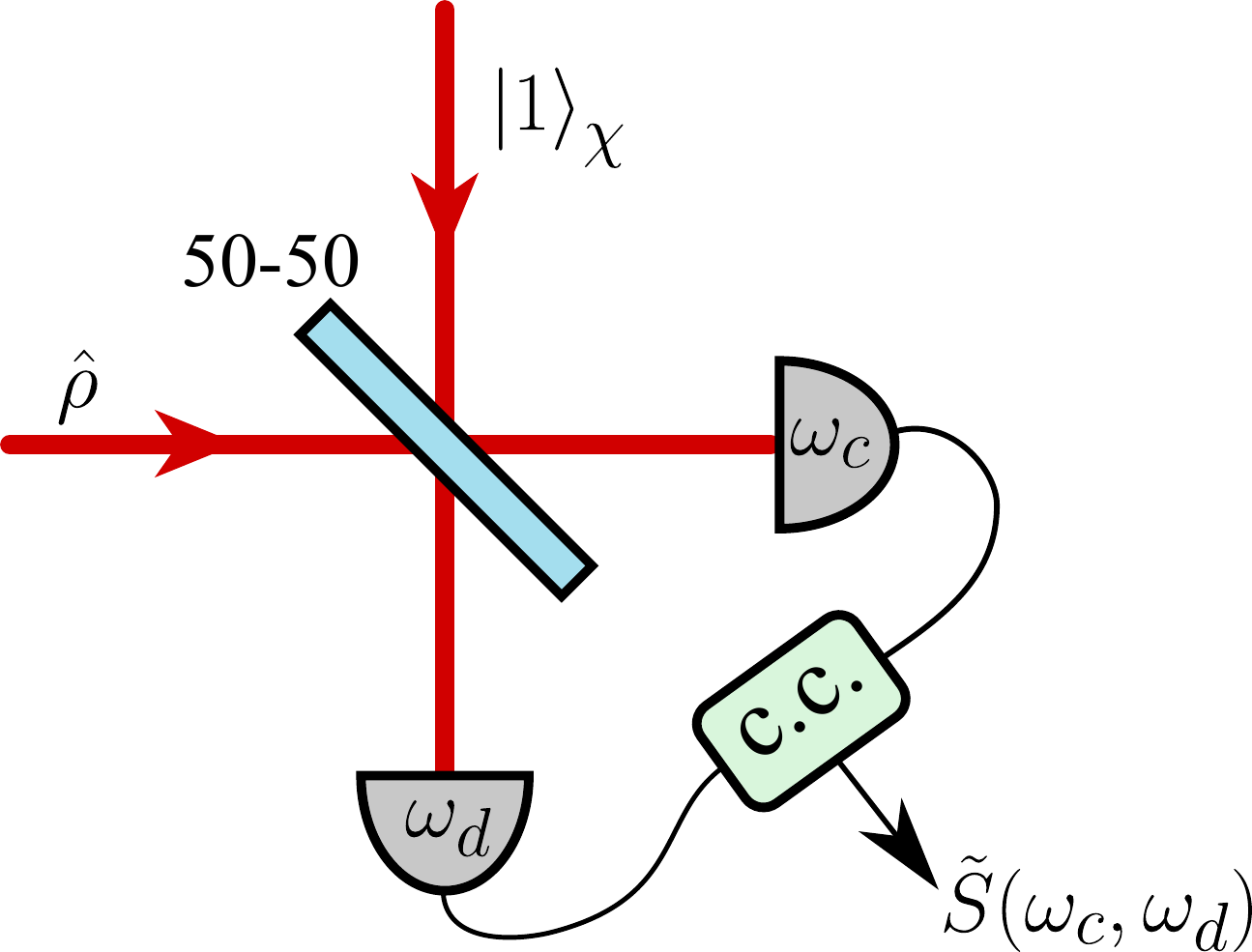}
	\centering
	\caption{General depiction of spectrally-resolved HOMI wherein the spectral coincidences between an unknown single photon described by a density matrix $\hat{\rho}$ and a reference photon in a pure state $\ket{1}_\chi$ are retrieved in the form of a two-dimension interferogram $\tilde{S}(\w_c,\w_d)$.}
	\label{homi-scheme}
\end{figure}
Here we employ spectrally-resolved detection at the output of the HOMI (as depicted in Fig. \ref{homi-scheme}), which significantly reduces the experimental complexity and data acquisition time required to determine the spectral density matrix elements. Spectrally resolved detection is achieved using time-resolved single photon spectrometers (TRSPS)\cite{davis:17}. 

Consider a train of single-photon reference pulses at input $b$ of the balanced beam splitter occupying a single mode, $\tilde{\chi}(\w)$, with spectral support that spans that of the unknown single-photon source
. If the unknown source emits single photons occupying the mode $\tilde{\psi}(\w)$ at input $a$ of the beam splitter, then the probability to register frequency-resolved coincidence detection at outputs $c(d)$ of the beam splitter with frequencies $\w_c(\w_d)$ is
\begin{equation}
S(\w_c,\w_d) \sim |\tilde{\psi}(\w_c)\tilde{\chi}(\w_d)-\tilde{\psi}(\w_d)\tilde{\chi}(\w_c)|^2.
\label{JointInterferogram}
\end{equation}

\begin{figure*}[t]
	\includegraphics[width=.85\linewidth]{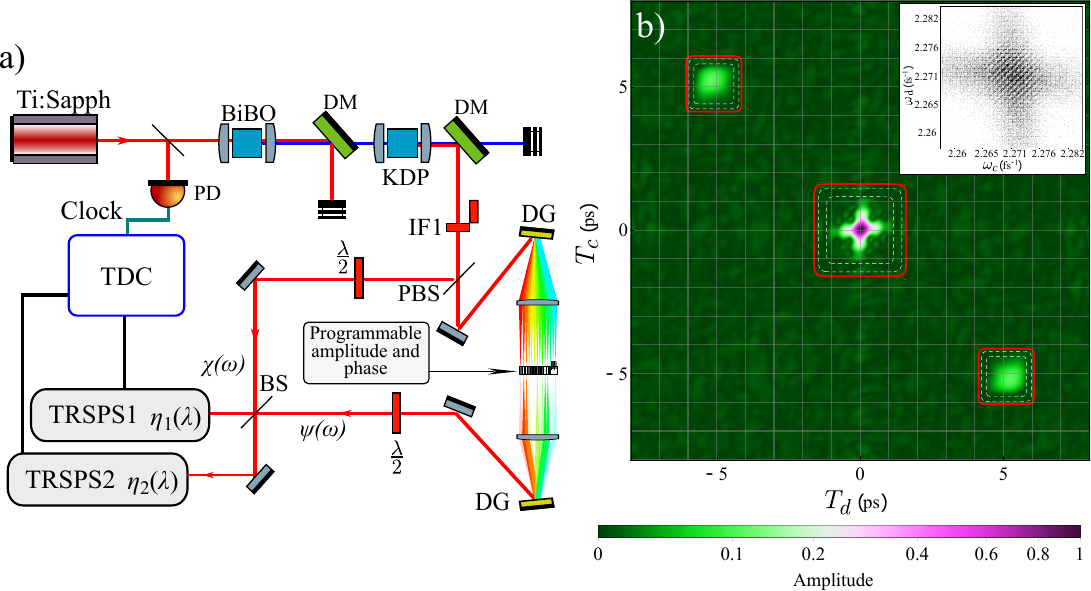}
	\centering
	\caption{a) Experimental scheme for pulse characterisation by spectrally-resolved Hong-Ou-Mandel interference using the idler photon as a reference. Note that the final beam splitter and the spectrometers are fibre-coupled. TDC: Time-to-digital converter. PD: photodiode. DM: dichroic mirror. DG: diffraction grating. (P)BS: (Polarising) beam splitter. b) Example 2-dimensional Fourier transform of two-photon spectral interferogram $S(\w_c,\w_d)$ (inset), giving the pseudo-temporal distribution $\bar{S}(T_c,T_d)$ in which the three peaks corresponding to the one unmodulated and two oscillatory terms are evident.}
	\label{scheme-and-fringes}
\end{figure*}

To extract the amplitude and phase of the density matrix $\rho(\w_1,\w_2)$, we will first make the assumption that the temporal duration of each pulse is much less than some time $\tau$, and hence the spectral intensity of neither pulse contains rapid changes in spectral amplitude. In particular, it allows us to apply delay $\tau$ in the reference pulse relative to the signal, such that $\chi(\w)\rightarrow \chi(\w)e^{i\w\tau}$, and be confident that the pulses do not overlap in time when passing through the beam splitter. The resulting spectral interference pattern (with introduced fringes) can then be written
\begin{align}
S(\w_c,&\w_d) =~|\tilde{\psi}(\w_c)|^2|\tilde{\chi}(\w_d)|^2+|\tilde{\chi}(\w_c)|^2|\tilde{\psi}(\w_d)|^2  \label{twodiminterferogram} \\ 
&+ \left( \tilde{\psi}(\w_c)\tilde{\psi}^*(\w_d)\tilde{\chi}(\w_d)\tilde{\chi}^*(\w_c)e^{i(\w_d-\w_c)\tau} + \textrm{c.c}\right) \nonumber 
\end{align}
Performing a 2D Fourier transform on the recovered interference pattern yields the pseudo-temporal intensity distribution $\bar{S}(T_c,T_d)$, which features three non-overlapping peaks separated by the delay $\tau$ (see Fig.\ref{scheme-and-fringes}b). The central peak corresponds to the first two terms, whilst the displaced peaks each correspond to one of the phase-sensitive terms. Since the peaks do not overlap in the pseudo-temporal domain, it is possible to multiply by some filter $f(T_c,T_d)$ to obtain the part of the distribution that is attributable only to one of the interference terms, $\bar{S}_f(T_c,T_d)$. Typically one would choose $F(T_c,T_d)$ to have a value close to unity in the region of pseudo-temporal space where one expects to find the interference term, and zero elsewhere. This filtering operation is depicted in  Fig. \ref{scheme-and-fringes}b, in which the contours correspond to the boundaries of the filters.
\begin{align}
\bar{S}_f(T_c,T_d)=&f(T_c,T_d)S(T_c,T_d) \\
=&\mathcal{F}\{\psi(\w_c)\psi^*(\w_d)\chi(\w_d)\chi^*(\w_c)e^{i(\w_d-\w_c)\tau}\}.\nonumber
\end{align}
From this point, it remains only to inverse-Fourier transform and divide by the known quantity $\chi(\w_d)\chi^*(\w_c)e^{i(\w_d-\w_c)\tau}$, after which the full complex quantity $\psi(\w)$ can be easily extracted, for example by matrix diagonalisation.

Where it cannot be assumed that the test pulse is in a pure state, this method generalises straightforwardly to the mixed-state formalism and allows full extraction of the spectral density matrix $\rho(\w_1,\w_2)$ without further measurements.

Since the final stage of the extraction is division by the spectral amplitude of the reference, it is necessary that the reference pulse completely covers the spectral support of the test signal. Spectral gaps in the reference photon will lead to the inability to characterise the phase of the signal at those frequencies. 

\section{Experimental scheme}

Our experimental scheme is depicted in Fig. \ref{scheme-and-fringes}a. A commercial Ti:Sapph femtosecond oscillator (Spectra-Physics Tsunami) delivers pulses with a bandwidth-limited duration of $100$ fs FWHM at $12.5$ ns intervals, corresponding to a spectrum centered at $830$ nm and a bandwidth of $10$ nm FWHM with a comb structure with a spacing of $80$ MHz. A beam pickoff at the output of the laser directs some light to a fast photodiode, which generates an electronic time-reference signal or `clock' that is used throughout the experiment. 

Second harmonic generation is achieved in a $1$ mm-long bismuth borate (BiBO) crystal, generating a $3$ nm FWHM spectrum at $415$ nm. Heralded single photons are generated by collinear, type-II SPDC in an $8$ mm-long potassium dihydrogen phosphate (KDP) crystal \cite{mosley:08}. The orthogonally-polarised signal and idler fields have, respectively, $12$ nm and $3$ nm bandwidths and degenerate 830 nm centre wavelengths. The two beams are separated at a polarising beam splitter (PBS). To achieve the experimental goals, both photons need to be recombined with a given temporal delay and their resulting interference pattern spectrally resolved. 
We use as a reference the broadband idler photon produced by the down-conversion whilst the narrowband signal photon is treated as the unknown state. Due to the strong correlations in photon number between the signal and idler arms, this approach greatly increases the probability of desired coincidences relative to the background two-photon events relative to an attenuated coherent state reference. However, this solution has the drawback of requiring the independent characterisation of the idler photon. For the purposes of this demonstration, the full spectral modefunction of the idler photon was determined by a direct reconstruction using the apparatus detailed in \cite{PRA:18,PRL:18}. However, if full characterisation of the reference photon is not possible, the scope of the measurement recovers the relative phase between the test photon and the signal (provided both have a common spectral support), which is still of considerable interest for mode-matching applications.

To test that our characterisation scheme can perform arbitrary mode reconstruction, we prepare ensembles of test pulses in a range of states by directing the signal photon to a fibre-coupled pulse shaper capable of performing arbitrary spectral phase operations, as well as spectral amplitude carving \cite{weiner:00}. The shaper consists of a standard $4$-f line built with a $2000$ lines/mm diffraction grating (Spectrogon) and $200$ mm focal length cylindrical lens (Thorlabs) with a $2$D phase mask (Hamamatsu SLM, $1272 \times 1024$ pixel mask) and is capable of achieving arbitrary spectral phase shaping with a resolution of $0.015$ nm/pixel and near-uniform spectral intensity transmission of $\approx 60$\%. Losses are equally distributed between the diffraction grating efficiency and the insertion losses in fibre coupling at the output of the device. 

The first experimental requirement is to ensure that the photons are indistinguishable in both the spatial and polarisation domains. The former was accomplished by implementing the interferometer in polarisation-maintaining single-mode fibre, using an evanescently coupled fibre beam splitter to close the interferometer. This ensured any photon events registering at the outputs correspond only to photons propagating in the fundamental spatial mode of the fibre.
Polarisation matching was achieved by coupling both signal and idler fiber into a polariser which output was monitored with a single-photon counting module. Transmission through the polarizer was maximized for the two down converted modes using half and quarter waveplates, thus verifying that the two photons were in the same polarisation mode.

\begin{figure}[t]
	\includegraphics[width=\linewidth]{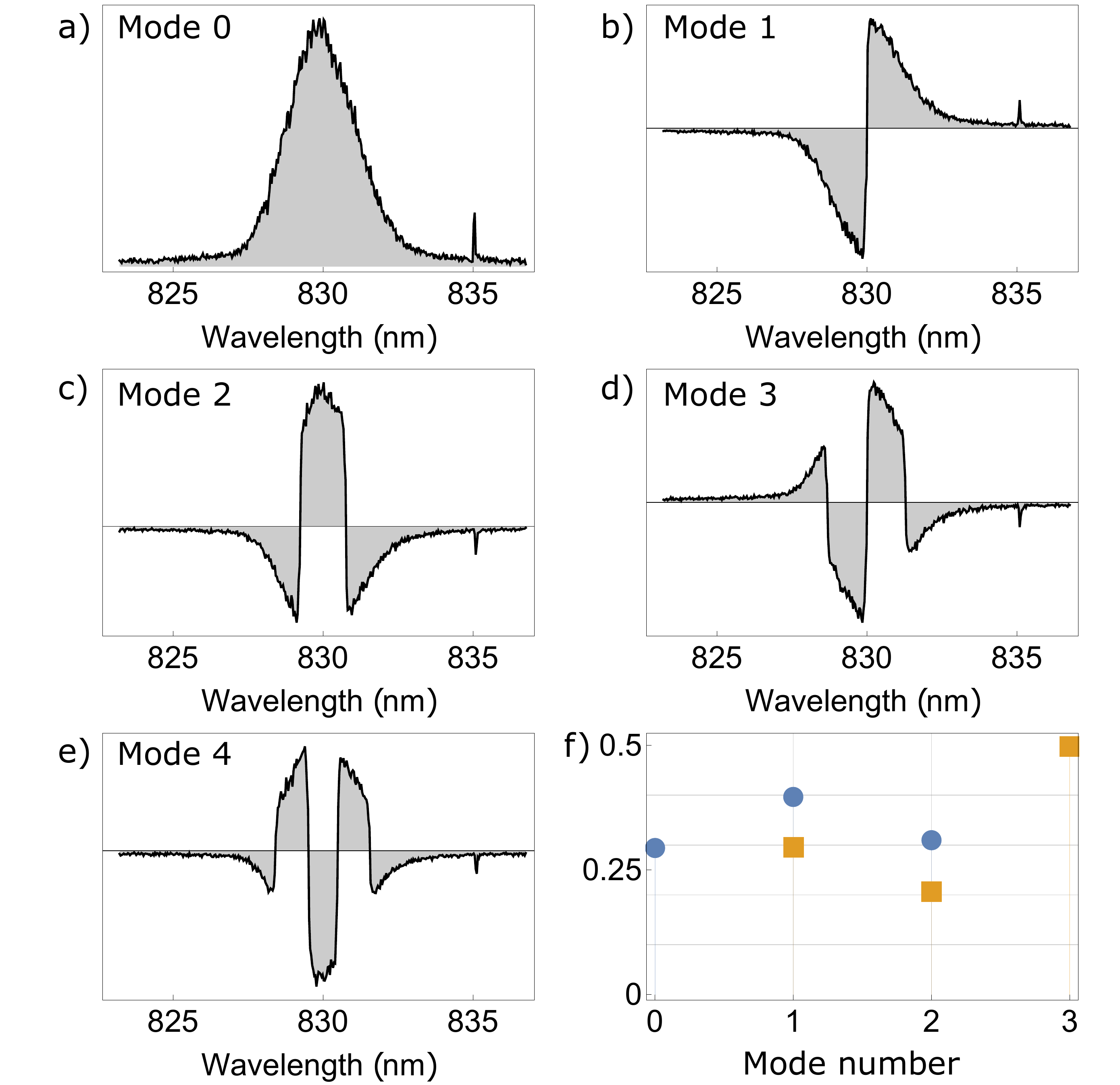}
	\centering
	\caption{a-e) Theoretical test modes obtained by pulse shaping. a) is the signal photon spectrum directly measured, whilst the remaining have been applied the $\pi$-phase jumps according to the setup of the pulse shaper. f) mode mixing ratios $P_i$ for mixed states A (dots) and B (square).}
	\label{modes}
\end{figure}

Another crucial step in the experimental preparation is to precisely set an appropriate value of $\tau$, which must be large enough for the terms in \eqref{twodiminterferogram} to be distinguishable in the pseudo-temporal domain, but small enough that the spectral fringes are sufficiently slow to be resolvable by the TRSPSs. A value in the range of several picoseconds was determined to be optimal, allowing the interference fringes to be clearly distinguished without loss of contrast from the spectral point-spread function of the TRSPSs. The two paths were matched using fast single photon detectors whilst fine tuning was achieved by scanning the linear spectral phase between the two path with the pulse shaper over a 30 ps window. An interference filter of 3 nm FWHM bandwidth was inserted in the broadband idler beam to maximize the spectral mode overlap. The delay that minimised the distinguishability of the two photons, and hence the number of coincidences, was taken to correspond to the point where $\tau=0$. The delay line was then moved to the setting where $\tau=5.5$ ps, the interference filter was removed, and the outputs of the interferometer routed into the two TRSPSs. 

After the two photons are separated by a PBS, they are routed into separate optical paths. The path of the signal mode passes through the pulse shaper, which also served to fine-tune the relative delay $\tau$ between the two arms by writing linear spectral phase onto the signal photon. The two photons are then recombined at a beam splitter, after which they are directed into time-of-flight spectrometers and undergo spectrally-resolved coincident detection with a resolution of $0.05$ nm\cite{davis:17}.

Using the pulse shaper, the signal photon was prepared in a variety of different modes to test the fidelity of the reconstruction. This approach is sufficient to show if the setup can reconstruct arbitrary modes for two reasons. First, the spectrum of the SPDC source was already known by direct spectrally resolved measurement. Second, no birefringent element could cause a difference in spectral phase between the two photons, which are degenerate in centre wavelength, and hence accumulate the same phase information until the interferometer is closed. Therefore, any extracted phase would result from what is imprinted by the pulse shaper, which allows comparison of the experimental results with the theoretical expectation.

We chose to prepare the single-photon pulses in a series of modes approximating the first five Hermite-Gauss modes \cite{brecht:15} by applying a $\pi$-phase shift across the points in the spectrum corresponding to the nodes of those states. This allowed a nontrivial and near-orthogonal basis of pseudo-Hermite-Gaussian modes to be prepared without the need for spectral amplitude carving, which introduces losses. Moreover, constant phase is challenging to extract using reference-free mode reconstruction methods, even using classical light. Note that the sharp phase shift signifies that the nematic crystals on the SLM mask change orientation by 180$^\circ$ over two adjacent pixels, resulting in severe diffraction and hence loss of amplitude. Hence a small amount of light is lost at these nodes (see Fig.\ref{modes}).

\begin{figure*}[t]
	\includegraphics[width=\linewidth]{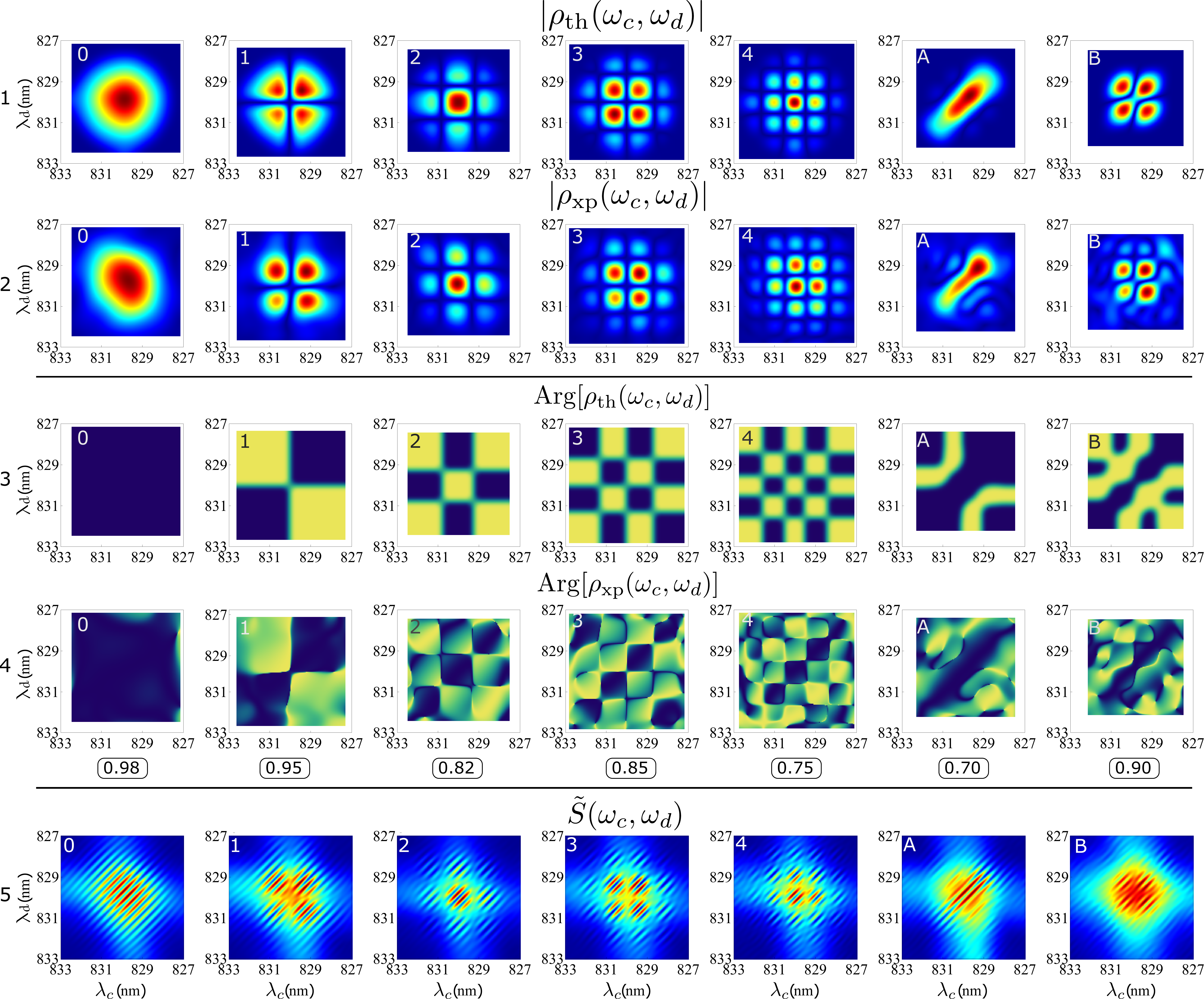}
	\centering
	\caption{Row 1-2: modulus of the theoretical and experimental density matrices for the five test pure states (0-4) and the two mixed states (A-B). Row 3-4: argument of the density matrices similar to the two top rows. Row 5: low-pass filtered interferograms. Below row 4: overlap between the theoretical and experimental matrices.}
	\label{allfigs}
\end{figure*}

\section{Results}
 The signal photon acquires a quadratic component of the spectral phase in propagation through 2 m of coupling fibre (approximately 90,000 fs$^2$).  In the case of the zero-order mode, this extra contribution was measured by the experiment. The pulse shaper then applied an additional component to the spectral phase to compensate for this, yielding a resultant flat spectral phase. Note that this value is consistent with the one measured in \cite{PRA:18,PRL:18}. In the subsequent cases, the measurement was taken with this quadratic spectral phase component compensated for.

The interferograms $S(\w_c,\w_d)$ are shown in the bottom row of Fig. \ref{allfigs}. Fourier filtration was used to remove high frequency noise of the detection (see Fig.\ref{homi-scheme}b). Isolating only the sideband, the density matrices were extracted using the procedure described above. The modulus of these, $|\rho(\w_1,\w_2)|$ as well as the phase $\phi(\w_1,\w_2)=\mbox{Arg}\{\rho(\w_1,\w_2)\}$ are presented in Fig. \ref{allfigs} (rows 2 and 4).

The boundaries of the regions of alternating phase are consistent with their predicted positions, indicating good agreement between the TRSPSs and the pulse-shaper calibration. In addition, a phase difference of approximately $\pi$ was measured between these regions, demonstrating the capability of this technique to characterise spectral phase.

Lastly, a mixed state was simulated by flickering the pulse shaper alternately between modes 0,1 and 2 (mixed mode A) and 1,2 and 3 (mixed mode B). This has the effect of simulating a mixed state described by a density operator comprising of a mixture of these three pure states. Interferograms for these two mixed states are also shown in Fig. \ref{allfigs} (bottom row), in which overlayed interference patterns from the different contributions are visible. The modulus and phase of the density matrices is shown in Fig.\ref{allfigs} (row 2 and 4). The suppression of the off-diagonal elements of the density matrix is clear, as the state begins to approximate a spectrally incoherent mixture for mode A, while that effect is less visible for mode B.

To further validate the reconstruction, since we know the relative spectral phase between the signal and idler path is flat (apart from the linear phase responsible for delay which is ignored during the reconstruction), we can construct theoretical density matrices $\rho_\textrm{th}$ from the measured spectra depicted in Fig.\ref{modes}. We normalize both theoretical and experimental matrices according to $\int \ud\w \ud\w' \modulus{\rho(\w,\w')}^2=1$, and we compute the overlap $\modulus{\int \ud\w \ud\w' \rho_\textrm{xp}^\ast(\w,\w') \rho_\textrm{th}(\w,\w')}$ which represents how similar the two matrices are. The values are shown at the bottom of row 4 in Fig.\ref{allfigs}. We find very good overlap for lower order pure state whilst the highest order reconstruction is poorer. The main reason for that decrease is not necessarily only a poor reconstruction. Since Fourier filtering is utilized to extract the density matrix, sharp variations in both amplitude and phase are smoothed out, whereas the theoretical matrices are constructed analytically and are much sharper. Consequently, the overlap quantity might not be the most accurate figure of merit, but it should provide a good lower bound. In fact, visually comparing the modulus of the matrices in Fig.\ref{allfigs} only reveals subtle differences, whereas the argument is more noisy due to the Fourier filtering and the algorithm that requires unwrapping of discontinuous phase.

More interestingly, we find that the mixed states show reasonable to good overlap with theory. The biggest difference arises again from the phase which is even more structured in that case, and hence harder to reconstruct through the algorithm. However, we note that the reconstruction is still accurate even though the modes were randomized during the acquisition.

\section{Conclusion}
In this work we proposed and demonstrated an externally-referenced technique to reconstruct the full spectral density matrix of an unknown single photon. The method has the advantages that it uses resolved measurements and hence does not require repeated reconfiguration in order to perform a reconstruction. Furthermore, there is no need to make the assumption of state purity- the full spectral density matrix of the photon is directly extracted from the interference pattern. Moreover, this method is potentially applicable to temporally mixed states, where each pure state has a different delay, since the information is contained along the diagonal in the pseudo-temporal Fourier domain. This approach may be of particular use when one photon from a well-characterised photon pair source is used to probe some unknown process and the other retained, in which case tomography on the state of the probe using this method can provide useful information about its evolution under the unknown process.


Future work on this technique may revisit the strategy of using a reference other than the twin photon from the same downconversion source. This strategy becomes practical when the spectrally-resolved detection efficiency is sufficiently high, and hence with advances in detector technology may become a feasible measurement.  Alternatively, a reference with sub-Poissian photon number statistics could be used to suppress the two-photon term. Ideally, the reference would be another single photon, with $g^{(2)}=0$. 

\section*{Acknowledgments}

	We are grateful to B. Brecht and M. Karpi\'{n}ski for fruitful discussions and insight on the results. This project has received funding from the European Union's Horizon 2020 research and innovation programme under Grant Agreement No. 665148, the United Kingdom Defense Science and Technology Laboratory (DSTL) under contract No. DSTLX-100092545, and the National Science Foundation under Grant No. 1620822. 


\bibliography{references}
	
\end{document}